\documentclass{article}
\usepackage{graphicx} 
\usepackage{graphicx}
\usepackage{epstopdf}
\usepackage{amssymb,amsmath,epsfig}
\usepackage{hyperref}
\usepackage{placeins}
\usepackage{subcaption}
\title{f(r) article}
\author{anam zahra}
\date{October 2024}
\begin{document}
\title{\bf Modeling Compact Objects in $f(R)$ Gravity: Application of Buchdahl-I Metric with Chaplygin Equation of State}
\author{
A. Zahra \thanks{anam.zahra@vsb.cz} \thanks{F2019265004@umt.edu.pk} ${}^{(a,b)}$, 
S. A. Mardan \thanks{syedalimardanazmi@yahoo.com} ${}^{(b)}$,
Muhammad Bilal Riaz 
\thanks{muhammad.bilal.riaz@vsb.cz} ${}^{~(a,c)}$,
\thanks{bilalsehole@gmail.com}\\
S. Saleem \thanks{saakhtar@kku.edu.sa} ${}^{~(d)}$,\\
${}^{a}$IT4Innovations, VSB – Technical University of Ostrava, \\Ostrava, Czech Republic,\\
${}^{b}$ Department of Mathematics,\\ University of the Management and Technology,\\
C-II, Johar Town, Lahore-54590, Pakistan.
\\${}^{c}$Jadara University Research Center, Jadara
University, Jordan,\\
${}^{d}$Department of Mathematics,\\ College of Science, King Khalid University, Abha, Saudi Arabia\\Center for Artificial Intelligence (CAI), King Khalid University,\\ Abha 61421, Saudi Arabia.
}
\date{}
\maketitle
\begin{abstract}
This paper investigates realistic anisotropic matter configurations for spherical symmetry in the framework of $f(R)$ gravity. The solutions obtained from Buchdahl-I metric are used to determine the behavior of PSR J0740+6620, PSR J0348+0432 and 4U 1608-52 with Starobinsky model. Analysis of physical parameters such as density, pressure, and anisotropy is illustrated through graphs, and the stability of compact objects is investigated by energy and causality conditions. We will also discuss the behavior of gravitational, hydrostatic and anisotropic forces, gravitational redshift and adiabatic index. At the theoretical and astrophysical scales, the graphical representations validate the practical and realistic $f(R)$ gravity models.

\end{abstract}
\section{Introduction}
Modified gravity theories extend general relativity (GR) to account for phenomena that GR alone cannot fully explain, including dark energy, dark matter, and the quest for a unified framework incorporating quantum mechanics. These modifications often involve changes to the Einstein-Hilbert action or the introduction of new fields, like scalar or vector fields, while reducing to GR under certain conditions. There are different theories of gravity such as $f(R)$, $f(R,T)$, $f(G)$ and $f(R,G)$ \cite{kausar2014dissipative}-\cite{nashed2021anisotropic}. Modified gravity offers alternatives to dark matter and dark energy by providing frameworks that aim to resolve galaxy rotation curves and the accelerated expansion of the universe, and they also serve as theoretical tools for testing gravity on cosmological and astrophysical scales \cite{{nojiri2007introduction}, {clifton2012modified}, {otiriou2010f}}. Capozziello and Laurentis \cite{capozziello2011extended} explored the fundamental concepts of \( f(R) \) gravity, emphasizing its role in bridging dynamical and conformal theories, as well as addressing the initial value problem. Cruz-Dombriz and Saez-Gomez \cite{de2012black} examined modified gravity theories, formulated an \( f(R) \) model to derive black hole solutions, and compared their findings with general relativity (GR) results. Bamba et al. \cite{bamba2013modified} investigated dark energy cosmologies within the framework of \( f(R) \) gravity, focusing on the implications of finite-time future singularities. Paul and Kalita \cite{paul2024solar} tests $f(R)$ gravity theory in the solar system, comparing acceleration due to gravity for planets, trans neptunian objects, centaurs, scattered disk objects, and oort cloud objects with Newtonian and modified Newtonian dynamics predictions.

The \( f(R,T) \) gravity theory extends GR by modifying its action to include both the Ricci scalar \( R \), which describes the spacetime curvature. It also incorporates the trace of the energy-momentum tensor \( T \), representing the fluid distribution. This generalization allows for a broader framework for describing gravitational interactions. The function $f(R,T)$ is typically divided into terms dependent on $R$ and $T$. The inclusion of $T$ allows this theory to account for matter effects directly in the gravitational dynamics, enabling potential explanations for cosmic acceleration and modifications to how matter behaves under gravity without invoking dark energy. The dependence on $T$ implies that matter and energy could influence the spacetime curvature in new ways compared to GR, potentially providing new insights into the interaction between geometry and matter fields \cite{harko2011f}.
The $f(T)$ gravity model is based on torsion rather than curvature, using the torsion scalar $T$ from teleparallel gravity. In this theory, the Weitzenböck connection, which has zero curvature but non-zero torsion, is used to describe gravitational effects. By allowing the action to be a function of torsion, $f(T)$ gravity provides an alternative explanation for cosmic acceleration, making it a contender for theories that aim to replace dark energy and explain inflation \cite{cai2016f}.

Anisotropic matter, which has directionally dependent properties such as pressure, is important in astrophysical contexts, especially when studying compact objects (COs) like neutron stars and black holes. In neutron stars, accounting for anisotropic pressure leads to more accurate models of their internal structure and stability, affecting mass-radius relations and the maximum mass they can support before collapse. In black hole formation, anisotropic matter can influence the dynamics of gravitational collapse and the development of singularities. In cosmology, anisotropic matter affects early universe evolution and the formation of cosmic structures. Additionally, anisotropic fluid models offer improved descriptions of matter under extreme gravitational conditions, enhancing predictions about dense objects and providing tests for the boundaries of GR and alternative gravity theories \cite{Herrera1997}-\cite{Ivanov202}. Abbas and Nazar \cite{abbas2018complexity} examined the complexity factor for anisotropic fluid configurations within the framework of $f(R)$ gravity. Their research aimed to investigate how this factor, which measures the structural complexity of the system, operates within the framework of modified gravity. The results offer insights into the dynamics of self-gravitational systems in the theory of $f(R)$. Abbas et al., \cite{abbas2015anisotropic} discussed the exact solution for anisotropic strange stars in $f(R)$ theory. With a cosmological constant, anisotropic compact stars can be examined in light of GR \cite{hossein2012anisotropic}. It is possible to discuss the compact structure while taking into account different forms of radial pressure and energy density \cite{ivanov2017analytical}. For compact stars, a singularity-free solution can be obtained in \cite{deb2017relativistic}. 
Panotopoulos \cite{panotopoulos2017strange} discussed strange stars in the $f(R)$ theory using the Palatini, focusing on the $R + \frac{R^2}{6M^2}$ and $R - \frac{\mu^4}{R}$ models. It finds that modified gravity's effects on strange stars can be negligible or ruled out. 
Silveira et al. \cite{Silveira2021} presented a model that extends Schwarzschild’s homogeneous star by introducing a density transition zone near the surface. The numerical integration of the modified TOV equations in \( f(R) = R + \lambda R^2 \) reveals configurations that are finite throughout. Depending on the parameters, objects with varying compactness-comparable to or exceeding those in GR are found, with some configurations approaching the Buchdahl limit.

The study of CO offers valuable theoretical insights into the properties of dense matter systems. Recent advances in theoretical modeling have facilitated the translation of abstract theoretical concepts into analytical solutions. A crucial factor in obtaining solutions to Einstein field equations (EFEs) is the selection of an appropriate metric potential. Delgaty and Lake \cite{1998} introduced specific metric ansatzes to obtain exact solutions to EFEs for static and spherically symmetric configurations of ideal fluids. These solutions play an important role in the estimation of the potential physical properties of CO.

To explore the attributes of CO, the Buchdahl-I metric \cite{1959} proves to be highly effective. Tamta and Fuloria \cite{tamta2022study} investigated anisotropic stellar structures using the Buchdahl metric potential and found that their models were stable over a range of parametric values. Similarly, Maurya et al. \cite{2019Maurya} studied the Buchdahl-I metric ansatz to analyze anisotropic CO.
Maurya et al. \cite{2020Maurya} studied the stability of stellar structures in $f(R,T)$ theory, which allows for non-conservation of energy-momentum. It suggests using the Buchdahl ansatz for metric potential to provide exact solution to the spherically symmetric EFEs with a perfect fluid.
 The research compares Buchdahl solutions in Einstein gravity to those in $f(R,T)$ gravity, revealing comparable behaviors while also noting certain advantageous characteristics in Einstein gravity. 
Kumar et al. \cite{kumar2021generalized} discuss the generalized Buchdahl model for compact stars in $f(R,T)$ gravity extends the traditional Buchdahl framework to accommodate the effects of the energy-momentum tensor alongside the Ricci scalar. This model provides a more comprehensive approach to studying the structure and stability of compact stellar objects within modified gravity theories. By incorporating both matter contributions and geometric effects, it helps analyze how these factors influence the equilibrium conditions of such stars. 
The most precise calculation of the maximum mass of CO has been obtained using Shapiro delay and current pulsar research. Based on this analysis, the millisecond pulsar PSR J0740+6620 has a mass of \( 2.14^{+0.1}_{-0.09} \ M_\odot \) \cite{cromartie2020relativistic}.
 It is one of the massive neutron star ever observed. 
PSR J0348+0432 is another millisecond pulsar in a binary system with a white dwarf companion and is notable for its mass, which is estimated at $2.01\ M_\odot$. This system is also significant because it has been used to test GR predictions under strong-field conditions. The pulsar emits gravitational waves as it orbits its companion, leading to orbital decay that is consistent with predictions from Einstein's theory. The high mass of PSR J0348+0432, combined with the effects of gravitational wave emission, provides stringent tests on alternative theories of gravity and the nature of compact stars. Observations of this system have supported the idea that neutron stars can support very high masses, further constraining the equation of state (EoS) of dense matter \cite{antoniadis2013massive}. The X-ray binary system 4U 1608-52 consists of a neutron star paired with a low-mass companion star. The neutron star in this system is an X-ray transient, meaning that it alternates between periods of quiescence and periods of high X-ray activity. 4U 1608-52 is especially notable for its role in thermonuclear X-ray bursts, which occur when matter accreted from the companion star ignites on the surface of the neutron star \cite{poutanen2014effect}.

The purpose of this paper is to study the impact of $f(R)$ gravity models and anisotropic pressure on the modeling of realistic COs. For three different observational datasets of stellar structures, numerous types of structural properties are analyzed: the density and pressure anisotropy distributions, TOV equation, energy conditions, causality condition, gravitational redshift, and adiabatic index. 

The conditions, along with $\rho$, $p_r$, and $p_t$, are essential for modeling a star with charged anisotropic matter. Figs. (\ref{fig:1})-(\ref{fig:7}) highlight the smooth geometry and the model's stability to describe realistic COs. These analytical results have been derived using recent mass and radius measurements of COs such as  PSR J0740+6620, PSR J0348+0432 and 4U 1608-52. The layout of the paper is given as: The next part gives a brief introduction to $f(R)$ gravity for anisotropic configurations in static spherical symmetry. Sect. 3 presents a discussion on several $f(R)$ gravity models, focusing on their physical feasibility. Sect. 4 evaluates the physical applicability of three well-established stellar structures. In Sect. 5, we will discuss the physical parameters of Chaplygin EoS.
Sect. 6 presents the stability analysis of PSR J0740+6620, 
PSR J0348+0432 and 4U 1608-52 . Finally, the key findings are summarized in the concluding section.
\section{Anisotropic Modified $f(R)$ Theory of Gravity}
This section introduces the Einstein-Hilbert action within the framework of \( f(R) \) gravity, expressed as 
\begin{equation}\label{1}
    S_{f(R)}=\frac{1}{2\kappa}\int d^4 x \sqrt{-g} f(R)+S_M,
\end{equation}  
where $\kappa$ is the coupling constant, $g$ is the determinant of the metric tensor, and $S_M$ denotes the action of the matter fields. The main idea behind this theory is to  extend GR by allowing the gravitational action to depend on the Ricci scalar rather than cosmological constant. After applying variation on $g$ in Eq. (\ref{1}), the field equation for $f(R)$ is obtained as
\begin{equation}\label{2}
    R_{ab}f_R -\frac{1}{2}f(R)g_{ab}+(g_{ab}\Box-\nabla_a\nabla_b)f_R=\kappa T_{ab}.
\end{equation}
Here, \( T_{ab} \) represents the energy-momentum tensor, \( \nabla_a \) denotes the covariant derivative operator, \( \Box \equiv \nabla^a\nabla_a \) corresponds to the d'Alembertian operator, and \( f_R \equiv \frac{df}{dR} \) signifies the derivative of \( f(R) \) with respect to the Ricci scalar \( R \). The Eq. (\ref{2}) can be represented as
\begin{equation}\label{3}
G_{ab} = \frac{\kappa}{f_{R}}(T_{ab}^{(D)}+T_{ab})\equiv T_{ab}^{(eff)},
\end{equation}
where $G_{ab}$ represent the Einstein tensor, $T_{ab}^{(D)}$ is effective stress-energy tensor. It can be stated as
\begin{equation}\label{4}
T_{ab}^{(D)}=\frac{1}{\kappa}\bigg\{\nabla _{a}\nabla _{b}f_{R}-\Box f_{R}g_{ab}+(f-Rf_{R})\frac{g_{ab}}{2}\bigg\}.
\end{equation}
We considered a spacetime that is static and spherically symmetric, expressed as
\begin{equation}\label{5}
ds^2=e^{\nu(r)}dt^{2}-r^2(d\theta^2+sin^2\theta d\phi^2)-e^{\lambda(r)}dr^{2}.
\end{equation}
The expression for the stress-energy tensor is defined as
\begin{equation}\label{6}
T_{ab}=(\rho +p_{t})u_{a}u_{b}-p_{t}g_{ab}+(p_{r}-p_{t})v_{a}v_{b},
\end{equation}
where $\rho$ is the fluid energy density, the radial and tangential pressure is denoted by $p_r$ and $p_t$ respectively. The four- vectors $u_{a}$ and $v_{a}$  satisfy the relations $u^{a}v_{a}=0$, $u_{a}u^{a}=1$ and $v_{a}v^{a}=-1$. 
From Eqs. (\ref{3}), (\ref{5}), and (\ref{6}), the $f(R)$ field equations can now be evaluated as
\begin{eqnarray}\nonumber
\rho &=&\frac{e^{-\lambda}}{2r^{2}}\Big(-2\lambda 'rf_{R}+2f_{R}-2f_{R}e^{\lambda}-2r^{2}f''_{R}+\lambda 'r^{2}f'_{R}-4rf'_{R}\\&+&r^{2}e^{\lambda}Rf_{R}-r^{2}fe^{\lambda}\Big),\label{7}\\
p_{r}&=&\frac{e^{-\lambda}}{2r^{2}}\Big(2e^{\lambda}f_{R}-2f_{R}-r^{2}f_{R}e^{\lambda}R+r^{2}e^{\lambda}f+2f_{R}r\nu '+4rf'_{R}\nonumber\\&+&\nu 'r^{2}f'_{R}\Big),\label{8}\\
p_{t}&=&\frac{e^{-\lambda}}{4r}\Big(-2f_{R}r\nu ''-\nu'^{2}rf_{R}-\nu '\lambda 'rf_{R}+2f_{R}\lambda '-2f_{R}\nu '+4f'_{R} \nonumber\\
&+&2\nu 'f'_{R}r+4rf''_{R}-2\lambda 'rf''_{R}+2e^{\lambda}rf-2e^{\lambda}rRf_{R}\Big).\label{9}
\end{eqnarray}
The associated Ricci scalar is given as
\begin{equation}\label{10}
    \frac{e^{-\lambda}}{2r^2}\Bigg(2r^2\nu''+r^2\nu'^2+4r\nu'-r^2\nu'\lambda'-4r\lambda'-4e^{\lambda}+4\Bigg).
\end{equation}
\section{Realistic Results for $f(R)$ Theory of Gravity}
We employ the modified Buchdahl-I metric to simulate anisotropic compact star formations in a realistic manner.
 The Buchdahl metric-I is given in \cite{kumar2021generalized} as
\begin{equation}\label{11}
    e^{\lambda(r)}=\frac{L(1+\chi r^2)}{L-\chi r^2},
\end{equation}
where $\chi$ and $L$ are constants. We considered a Starobinsky model presented in \cite{starobinsky1980new} for the Ricci scalar as
\begin{equation}\label{12}
    f(R)=R+\beta(R^2),
\end{equation}
where $\beta$ is a constant. The Starobinsky model is a theory of the early universe that explains how space rapidly expanded right after the Big Bang. It helps to match observations of the universe’s structure and the cosmic microwave background.
The $f(R)$ field equations are derived by using the Buchdahl-I metric given in Eq. (\ref{11}) and the Starobinsky model presented in Eq. (\ref{12}). By using Eqs. (\ref{10}), (\ref{11}), and (\ref{12}) into Eq. (\ref{7}), we obtained 
\begin{equation}\label{13}
\rho=\frac{-(1 + L) \alpha_1 \chi 
 }{L^3 (1 + r^2 \chi)^9}.
\end{equation}
where
\begin{eqnarray}\nonumber
   \alpha_1&=&  
    4 \beta \chi^2 \Bigg( 324 + 1296 r^2 \chi + 5913 r^4 \chi^2+3597 r^6 \chi^3 +1433 r^8 \chi^4 + 247 r^{10} \chi^5  \\\nonumber&+& 30 r^{12} \chi^6 \Bigg) 
    - L^2 \Bigg( 
        6 + (48 + 39 r^2 - 72 \beta) \chi + \Big(109 r^4 + 144 \beta - 2 r^2 (91\\\nonumber&+& 144 \beta)\Big) \chi^2  + 3 \Big(57 r^6 + 5584 r^2 \beta- 2 r^4 (139 + 79 \beta)\Big) \chi^3+ \Big(165 r^8 + 9612 r^4 \beta \\\nonumber&-& 2 r^6 (460 + 213 \beta)\Big) \chi^4  + r^6 \Big(101 r^4 + 3628 \beta - 4 r^2 (73 + 59 \beta)\Big) \chi^5 + r^8 \Big(39 r^4 \\\nonumber&+& r^2 (42 -84 \beta) + 412 \beta\Big) \chi^6  + r^{10} \Big(9 r^4 + r^2 (22 - 18 \beta) + 20 \beta\Big) \chi^7 + r^{12} (4 + r^2) \\\nonumber&&(r^2 - 2 \beta) \chi^8 \Bigg) + 2 L \chi \Bigg( 
        -(1 + r^2 \chi)^3  \Big( 54 + 129 r^2 \chi + 312 r^4 \chi^2 + 65 r^6 \chi^3 + 12 r^8 \chi^4 \Big) \\\nonumber&+&\beta \Big( 36 + 144 (4 + r^2) \chi  + 3 r^2 (-1928 + 79 r^2) \chi^2  + 3 r^4 (2340 + 71 r^2) \chi^3\\\nonumber&+& 2 r^6 (2690 + 59 r^2) \chi^4 +14 r^8 (190 + 3 r^2) \chi^5+ r^{10} (484 + 9 r^2) \chi^6 + r^{12} (64 + r^2) \chi^7 \Big) 
    \Bigg) 
\end{eqnarray}
The modified Chaplygin EoS is given in \cite{bhattacharjee2024maximum} as
\begin{equation}\label{14}
    p_r=H\rho-\frac{S}{\rho},
\end{equation}
where $H$ and $S$ are any positive constants. Using Eq. (\ref{13}) and Eq. (\ref{14}), we obtain the $p_r$ which is given as
\begin{eqnarray} \nonumber
p_r &=& 
- L^6 S (1 + r^2 \chi)^{18}- 
 \frac{H (1 + L) \chi\alpha_1}
 {
  L^3 (1 + r^2 \chi)^9}\label{15}.
\end{eqnarray}

Similarly, using Eqs. (\ref{10}), (\ref{11}), and (\ref{12}), we derive an equation of $p_t$ from Eq. (\ref{9}) which is given in the Appendix.
The anisotropy factor $\Delta$ is the difference between $p_t$ and $p_r$. 
The mass for self-gravitating COs is defined as
\begin{equation}\label{18}
m(r)=4\pi \int_{0}^{R} \rho r^2 dr.
\end{equation}
\section{Junction Conditions for $f(R)$ Theory}
In this section, we will match the interior and exterior spacetime at the boundary surface. This process establishes a seamless transition between the two regions. The external metric is represented as
\begin{equation}\label{19}
    ds^2 = - f(r) \, dt^2 + \frac{1}{f(r)} \, dr^2 + r^2 \left( d\theta^2 + \sin^2\theta \, d\phi^2 \right),
\end{equation}
where \( f(r) \) is represented as
\begin{equation}\nonumber
    f(r) = 1 - 2u.
\end{equation}
The following represent the junction conditions applied when matching two metric potentials.
\begin{equation}\label{20}
   e^{\nu}= e^{-\lambda}=1-2u,
\end{equation}
In this case, the star's compactness is represented by \( u = \frac{M}{R} \). At the boundary surface, the \( p_r \), is set to zero as
\begin{equation}\label{21}
    p_r(R)=0.
\end{equation}
Goswami et al., \cite{Goswami} and Clifton et al., \cite{15*} presented the following junction conditions for \(f(R) \) theories based on boundary continuity conditions.
\\
The metric components must be continuous on the boundary surface \(\Sigma\). 
\begin{equation}
  \left[ g_{ab} \right]^{+}_{-} = 0.
\end{equation}
where $g_{ab}$ represents the components of the metric tensor in the interior $(-)$ and exterior $(+)$ regions. For the given line element, $e^{\nu(r)}$ and $e^{\lambda(r)}$ should match $r=r_{\Sigma}$.
\\
The extrinsic curvature of the boundary surface \(\Sigma\) must also be continuous
 \begin{equation}
     f,_R \big[{K}_{ij}\big]^{+}_{-} = 0.
\end{equation}
\\
In \(f(R)\) gravity, the Ricci scalar is a dynamical variable and must be continuous across the boundary defined as
\begin{equation}
    \left[ R \right]^{+}_{-} = 0.
\end{equation}
\\
The trace of the extrinsic curvature, \( K \), remains consistent across the boundary surface. This ensures a smooth embedding of the hypersurface in spacetime.
\begin{equation}
    \left[ K \right]^{+}_{-} = 0.
\end{equation}
It is essential to take into account that the additional conditions listed above apply to $f(R)$ theories with a nonlinear function $f$. These additional conditions, which impose some significant restrictions on spacetimes, are required for the field equations to remain continuous across the matching surface. The value of $\chi$ and $S$ can be obtained from Eqs. (\ref{11}), (\ref{13}), (\ref{14}), and (\ref{20}) as
\begin{equation}
   \chi= \frac{2 L u}{r^2 (1 + L - 2 L u)},
\end{equation}
and
\begin{eqnarray}
S&=&\frac{1}{\Big( -L^6 (1 + r^2 \chi)^{18} \Big)^{1/3}} (-1)^{1/3} H^{1/3} (1 + L)^{2/3} \chi^{2/3} \Big( 
   -4 \beta \chi^2 \Big( 324\nonumber \\
   & +& 1296 r^2 \chi + 5913 r^4 \chi^2 + 3597 r^6 \chi^3 + 1433 r^8 \chi^4 + 247 r^{10} \chi^5 + 30 r^{12} \chi^6 \Big) \nonumber \\
   &+& L^2 \Big( 6 + (48 + 39 r^2 - 72 \beta) \chi + \Big( 109 r^4 + 144 \beta - 2 r^2 (91 + 144 \beta) \Big) \chi^2 \nonumber \\
   &+& 3 \Big( 57 r^6 + 5584 r^2 \beta - 2 r^4 (139 + 79 \beta) \Big) \chi^3 \nonumber \\
   &+& \Big( 165 r^8 + 9612 r^4 \beta - 2 r^6 (460 + 213 \beta) \Big) \chi^4 + r^6 \Big( 101 r^4 + 3628 \beta \nonumber \\
   &-& 4 r^2 (73 + 59 \beta) \Big) \chi^5+ r^8 \Big( 39 r^4 + r^2 (42 - 84 \beta) + 412 \beta \Big) \chi^6 \nonumber \\
   &+& r^{10} \Big( 9 r^4 + r^2 (22 - 18 \beta) + 20 \beta \Big) \chi^7+ r^{12} (4 + r^2) (r^2 - 2 \beta) \chi^8 \Big)\nonumber \\
   & -& 2 L \chi \Big( -(1 + r^2 \chi)^3 \Big( 54 + 129 r^2 \chi + 312 r^4 \chi^2 + 65 r^6 \chi^3 + 12 r^8 \chi^4 \Big) \nonumber \\
   &+& \beta \Big( 36 + 144 (4 + r^2) \chi+ 3 r^2 (-1928 + 79 r^2) \chi^2 + 3 r^4 (2340 + 71 r^2) \chi^3 \nonumber \\
   &+& 2 r^6 (2690 + 59 r^2) \chi^4+ 14 r^8 (190 + 3 r^2) \chi^5+ r^{10} (484 + 9 r^2) \chi^6 \nonumber \\
   &+& r^{12} (64 + r^2) \chi^7 \Big) \Big) 
\Big)^{2/3}.
\end{eqnarray}
\section{Physical Bounds on the Parameter of Chaplygin EoS}
A physically realistic model requires a positive $\rho$ and $p_r$ is finite at the centre when $r$ is zero. Eq. (\ref{13}) can be utilized to express the center energy density as 
 \begin{eqnarray}
\rho_0&=&\frac{1}{L^3}\times(1 + L) \chi \Big( -1296 \beta \chi^2 + L^2 \Big( 6 + (48 - 72 \beta) \chi + 144 \beta \chi^2 \Big)\nonumber\\ &-& 2 L \chi \Big( -54 + \beta (36 + 576 \chi) \Big) \Big).
\end{eqnarray}
By using Eq. (\ref{16}) the expression for central radial pressure is obtained as
\begin{eqnarray}
p_r(0)&=&\Bigg(L^3 (1 + L) \chi \Big( -1296 \beta \chi^2 + L^2 \Big( 6 + (48 - 72 \beta) \chi + 144 \beta \chi^2 \Big)\nonumber\\ &-& 2 L \chi \Big( -54 + \beta (36 + 576 \chi) \Big) \Big)\Bigg)^{-1}\times-\Bigg(L^6 S + H (1 + L)^2 \chi^2 \Big( 1296 \beta \chi^2 \nonumber\\ &-& L^2 \Big( 6 + (48 - 72 \beta) \chi + 144 \beta \chi^2 \Big)+ 2 L \chi \Big( -54 + \beta (36 + 576 \chi) \Big) \Big)^2\Bigg).\nonumber\\
\end{eqnarray}
The positivity of the central density is ensured when the values of $\chi$ and $L$ both are positive. 
\section{Physical aspects of $f(R)$ Theory}
The Starobinsky model in $f(R)$ theory is examined in the following section with the intent to describe certain physical characteristics of compact star interiors. We will use the Starobinsky model to examine the evolution of $\rho$, $p_r$, $p_t$, the hydrostatic equilibrium (TOV) equation, and energy conditions. We employ three different configurations of stellar objects, i.e., PSR J0740+6620, PSR J0348+04328, and 4U 1608-52 with masses $2.14^{+0.1}_{-0.09}\ M_\odot$, $2.01\pm0.04 \ M_\odot$, and $1.74M\ M_\odot$, respectively. We used the Starobinsky model and the Buchdahl metric-I to calculate the values of $\rho$, $p_r$, and $p_t$. The $f(R)$ field equations will be used to examine the various stability features associated with compact star structures.

In this paper, for the stability analysis of stellar structure of COs, we will use the causality conditions, gravitational redshift, adiabatic index. To investigate equilibrium conditions, the analysis will focus on how hydrostatic, gravitational, and anisotropic forces influence the system. Additionally, the internal structure of the stellar objects will be examined by generating plots to visualize and understand the behavior of these forces within the stars.
\subsection{Energy density, pressure and anisotropy}
\begin{figure}
\centering
\includegraphics[width=0.75\linewidth]{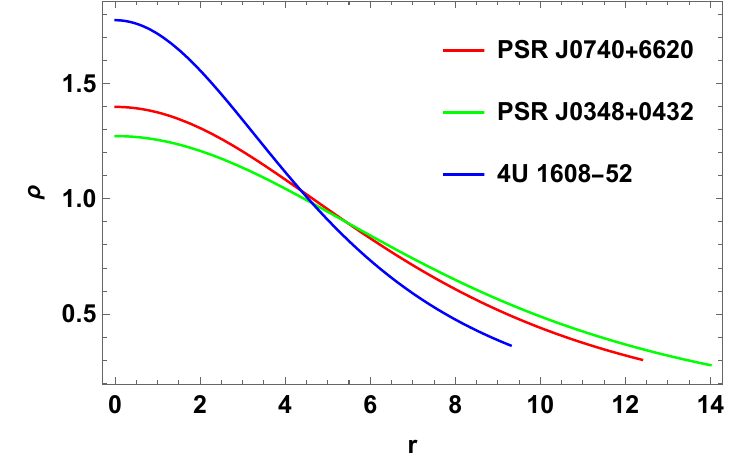}
\caption{The plot of energy density against $r$.}
\label{fig:1}
\end{figure}

\begin{figure}
\centering
\includegraphics[width=0.75\linewidth]{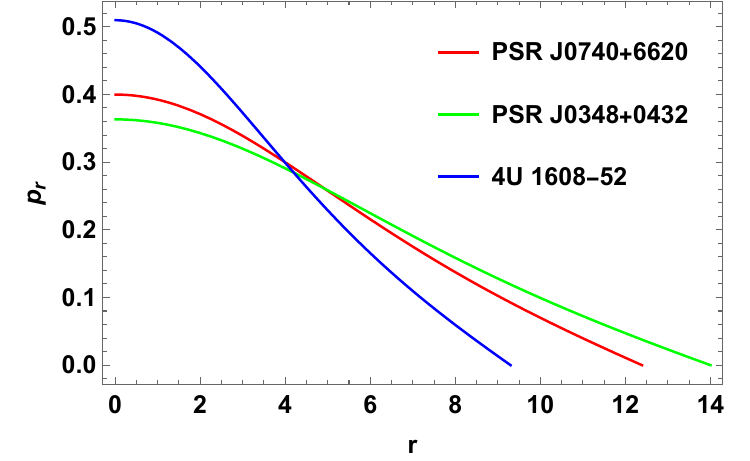}
\caption{The plot of radial pressure against $r$.}
\label{fig:2}
\end{figure}

\begin{figure}
\centering\includegraphics[width=0.75\linewidth]{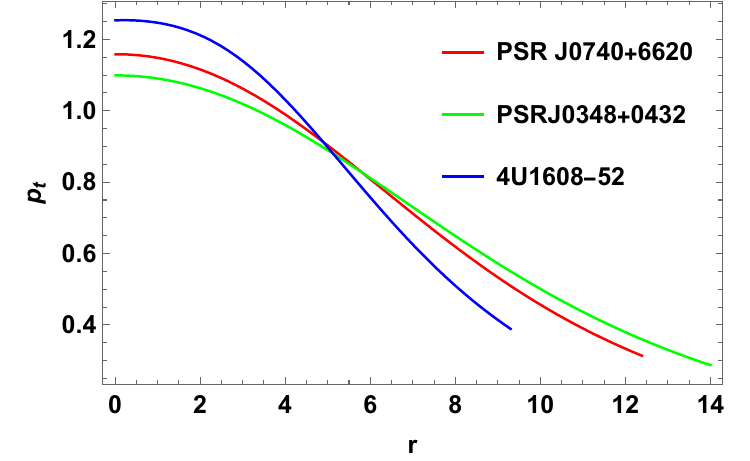}
\caption{The plot of transverse pressure against $r$.}
\label{fig:3}
\end{figure}

\begin{figure}
\centering
\includegraphics[width=0.75\linewidth]{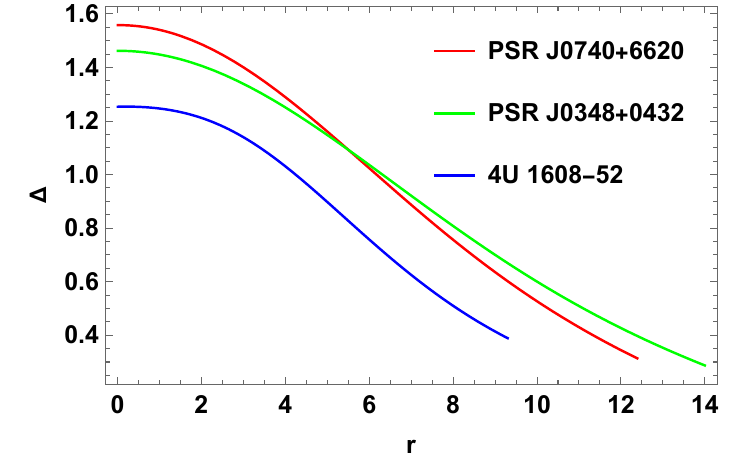}
\caption{The plot of anisotropy against $r$.}
\label{fig:4}
\end{figure}

\begin{figure}
\centering
\includegraphics[width=0.75\linewidth]{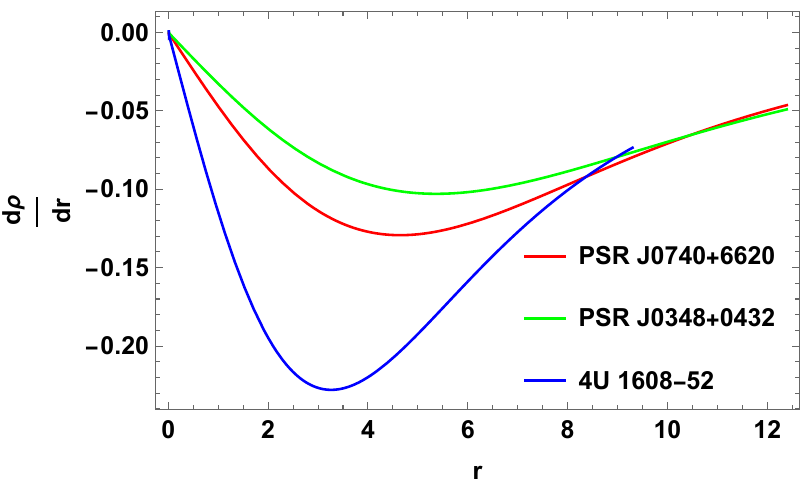}
\caption{The plot of $d\rho/dr$ against $r$.}
\label{fig:5}
\end{figure}

\begin{figure}
\centering
\includegraphics[width=0.75\linewidth]{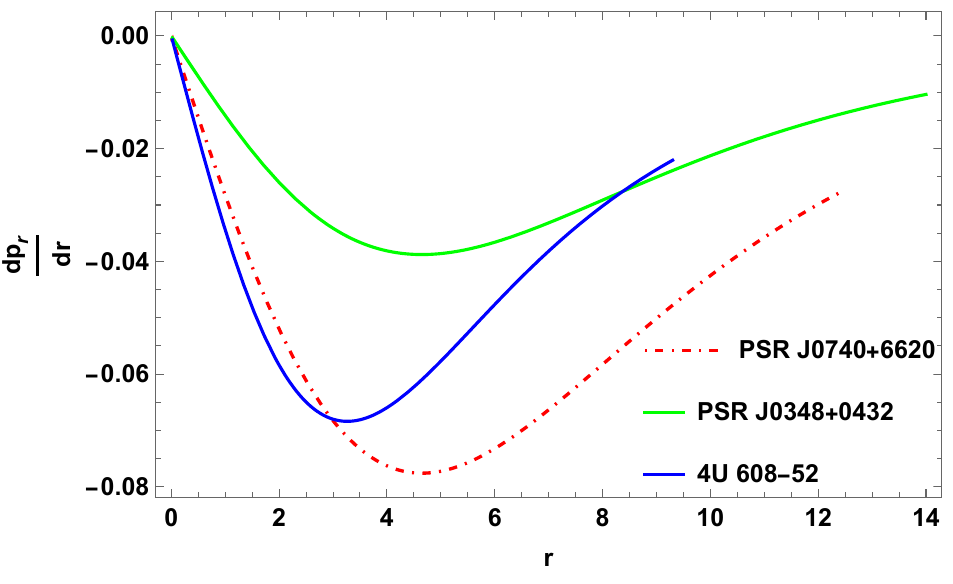}
\caption{The plot of $dp_r/dr$ against $r$.}
\label{fig:6}
\end{figure}

\begin{figure}
\centering
\includegraphics[width=0.75\linewidth]{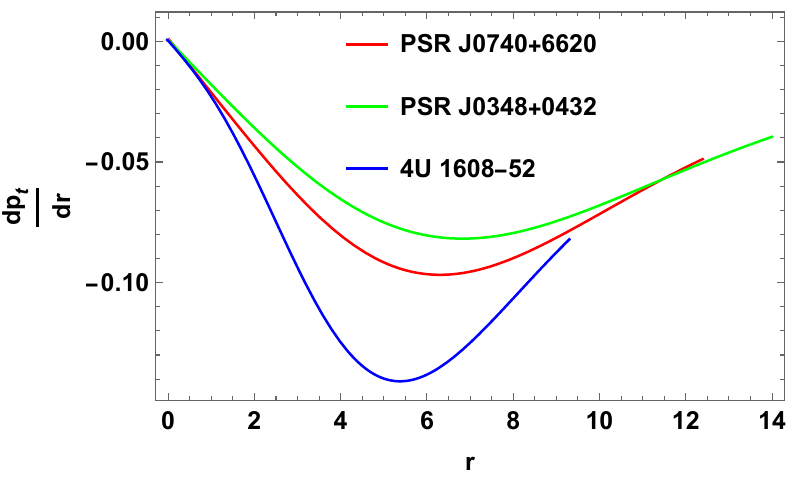}
\caption{The plot of $dp_t/dr$ against $r$.}
\label{fig:7}
\end{figure}

This section will demonstrate the characteristics of $\rho$, $p_r$, and $p_t$ with the $f(R)$ Starobinsky model. Fig. (\ref{fig:1}) illustrates the positive behavior of $\rho$. As the radius values increase, $\rho$ decreases, indicating a strong compactness in the star core. This suggests that the $f(R)$ model we have selected will yield positive results in the outer regime of the core.  Figs. (\ref{fig:2}) and (\ref{fig:3}) present the variation of $p_r$ and $p_t$. The difference $p_t-p_r$ is directly proportional to the quantity $\Delta$. The positive value of $\Delta$ signifies the positive value of $p_t-p_r$. However, if $\Delta$ is less than zero, the pressure resulting will be directed inward.  The behavior of $\Delta$ is presented in Fig. (\ref{fig:4}). The analysis will also focus on the dominant behavior of the stellar matter and the anisotropic pressure at the center using the $f(R)$ gravity model. Changes in the $\rho$, as well as $p_r$ and $p_t$, will be illustrated in the corresponding Figs. (\ref{fig:5}), (\ref{fig:6}), and (\ref{fig:7}). In the Starobinsky model, the study focuses on the $\rho$, $p_r$, and $p_t$ as functions of the radial distance from the center of a CO. It is observed that all three quantities decrease with increasing radial distance, as indicated by the negative derivatives: $\frac{d\rho}{dr}<0$, $\frac{dp_r}{dr}<0$, and $\frac{dp_t}{dr}<0$. This trend suggests that the energy density and pressures are higher near the center of the object and diminish outward.

At the center, where \(r=0\), the derivatives \(\frac{d\rho}{dr}\) and \(\frac{dp_r}{dr}\) become zero, indicating that there is no change in $\rho$, $p_r$, and $p_t$ at this point. This condition implies that the $\rho$, $p_r$, and $p_t$ reach their maximum values at the center before decreasing as one moves outward.

These characteristics are significant for understanding the structure and stability of COs like neutron stars and black holes. The decreasing profiles of energy density and pressure can influence the objects' evolution and stability against gravitational collapse, providing insights into their surface characteristics and overall behavior.
\FloatBarrier
\FloatBarrier
\FloatBarrier
\FloatBarrier
\FloatBarrier
\subsection{Energy conditions}
\begin{figure}[ht]
    \centering
\begin{subfigure}[b]{0.5\textwidth}
  \centering
  \captionsetup{justification=centering}
\includegraphics[width=0.75\linewidth]{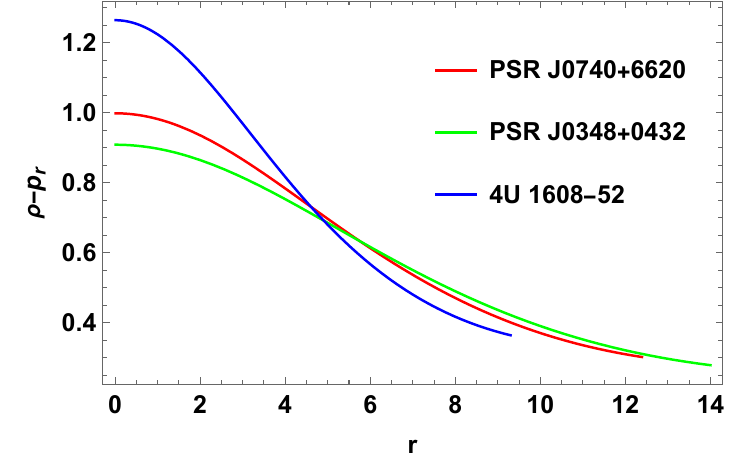}
\caption{}
\label{fig:subfig8a}
\end{subfigure}%
\hfill
\begin{subfigure}[b]{0.5\textwidth}
  \centering   
  \captionsetup{justification=centering}
\includegraphics[width=0.75\linewidth]{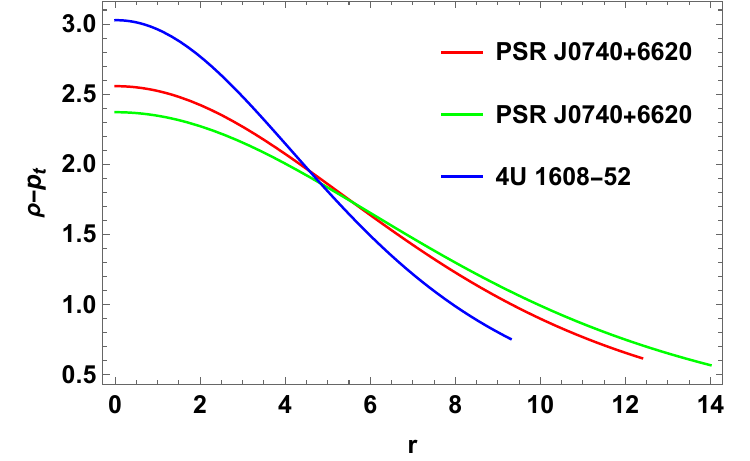}
\caption{}
\label{fig:subfig8b}
\end{subfigure}%
\hfill
\begin{subfigure}[b]{0.5\textwidth}
  \centering   
  \captionsetup{justification=centering}
\includegraphics[width=0.75\linewidth]{densitypr.pdf}
\caption{}
\label{fig:subfig8c}
\end{subfigure}%
\hfill
\begin{subfigure}[b]{0.5\textwidth}
  \centering   
  \captionsetup{justification=centering}
\includegraphics[width=0.75\linewidth]{densitypt.pdf}
\caption{}
\label{fig:subfig8d}
\end{subfigure}%
\hfill
\begin{subfigure}[b]{0.5\textwidth}
  \centering   
  \captionsetup{justification=centering}
\includegraphics[width=0.75\linewidth]{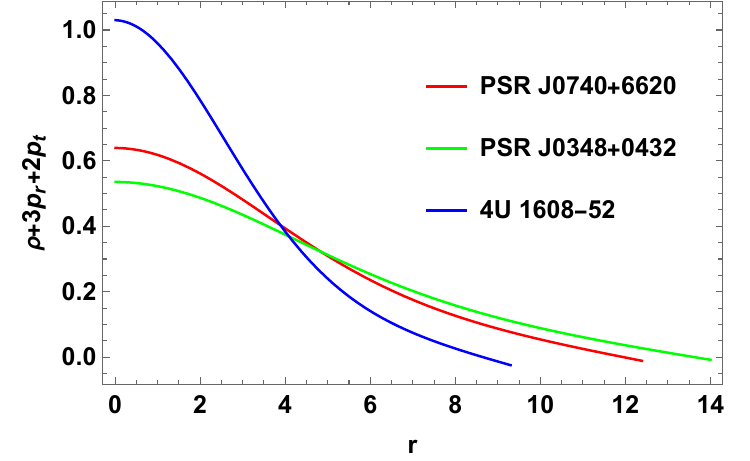}
\caption{}
\label{fig:subfig8e}
\end{subfigure}%
\caption{The plot of energy conditions against $r$.}
\label{fig:8}
\end{figure}
\subsection{Hydrostatic equilibrium (TOV) equation}
\begin{figure}[ht]
    \centering
\begin{subfigure}[b]{0.75\textwidth}
  \centering
  \captionsetup{justification=centering}
\includegraphics[width=0.75\linewidth]{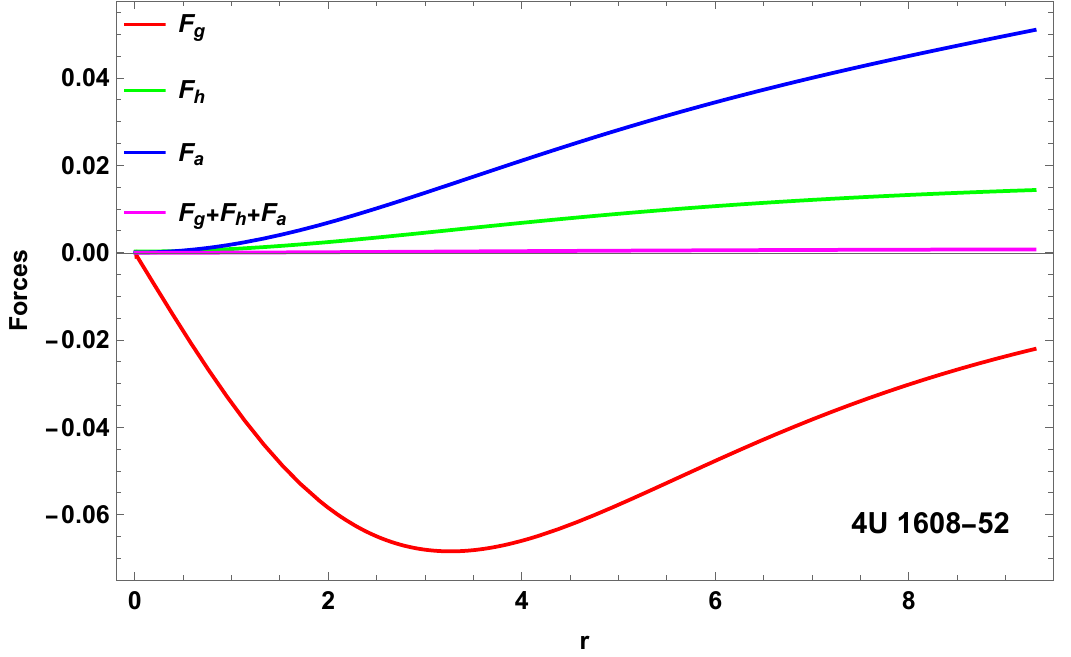}
\caption{}
\label{fig:subfig9a}
\end{subfigure}
\hfill
\begin{subfigure}[b]{0.75\textwidth}
  \centering   
  \captionsetup{justification=centering}
\includegraphics[width=0.75\linewidth]{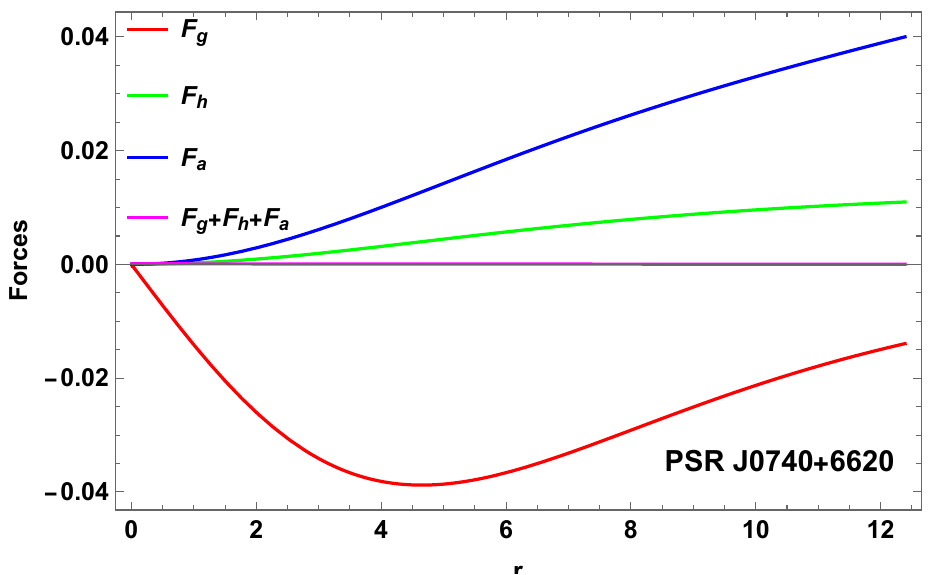}
\caption{}
\label{fig:subfig9b}
\end{subfigure}
\hfill
\begin{subfigure}[b]{0.75\textwidth}
  \centering   
  \captionsetup{justification=centering}
\includegraphics[width=0.75\linewidth]{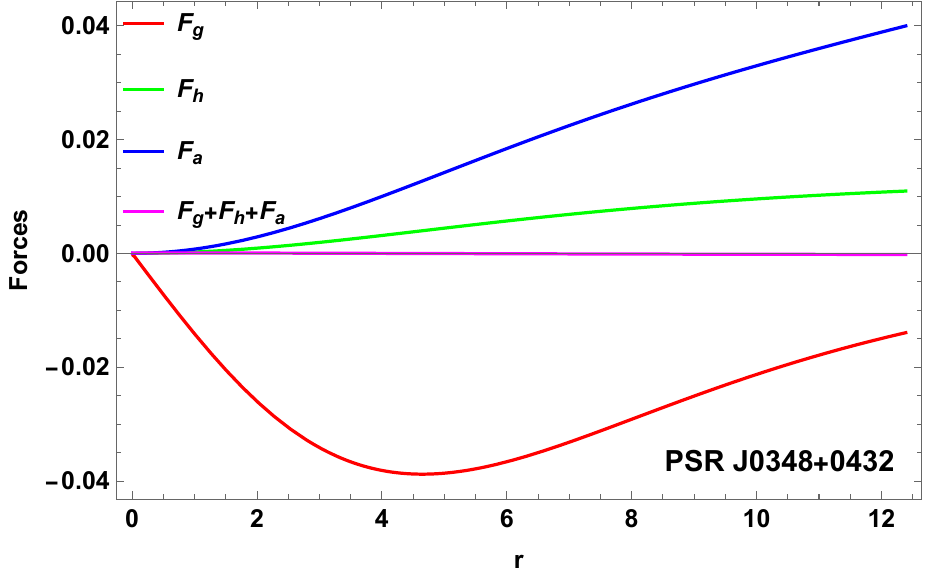}
\caption{}
\label{fig:subfig9c}
\end{subfigure}%
\caption{The plot of forces against r.}
\label{fig:9}
\end{figure}

The investigation of TOV equation is necessary to evaluate the stability of the model under different forces. The stability of charged anisotropic CO is influenced by the hydrostatic force (\( F_h \), the electrostatic force (\( F_e \), the gravitational force (\( F_g \), and the anisotropic force (\( F_a \). The model should be in equilibrium when all of these elements are taken into account. To examine stability, the following form of the generalized hydrostatic equilibrium condition, as established by Tolman \cite{1939Tolman} and Oppenheimer \cite{1939Oppen}, is employed.
The TOV is given as 
\begin{equation}\label{TOV*}
    -\nu^{\prime}(\rho+p_r)-\frac{dp_r}{dr}+\frac{2\Delta}{r}=0.
\end{equation}
The forces expressed as
\begin{eqnarray}
    F_g&=& -\nu^{\prime}(\rho+p_r),\label{fg}\\
    F_h&=&-\frac{dp_r}{dr},\label{fh}\\
    F_a&=&\frac{2\Delta}{r}.\label{fa}
\end{eqnarray}
It is possible to compute the expressions of Eqs. (\ref{fg})–(\ref{fa}) by utilizing Eqs. (\ref{13}), (\ref{15}), and (\ref{16}). Figs. \ref{fig:subfig9a}, \ref{fig:subfig9b}, and \ref{fig:subfig9c} provide a visual representation of the equilibrium circumstances of the model. 
\FloatBarrier
\subsection{Causality conditions}
\begin{figure}
    \centering
    \includegraphics[width=0.75\linewidth]{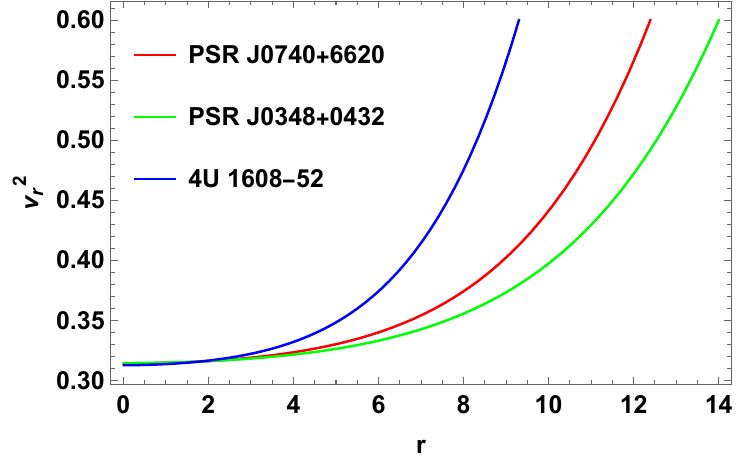}
    \caption{The plot of radial speed of sound against $r$.}
    \label{fig:10}
\end{figure}

\begin{figure}
    \centering
    \includegraphics[width=0.75\linewidth]{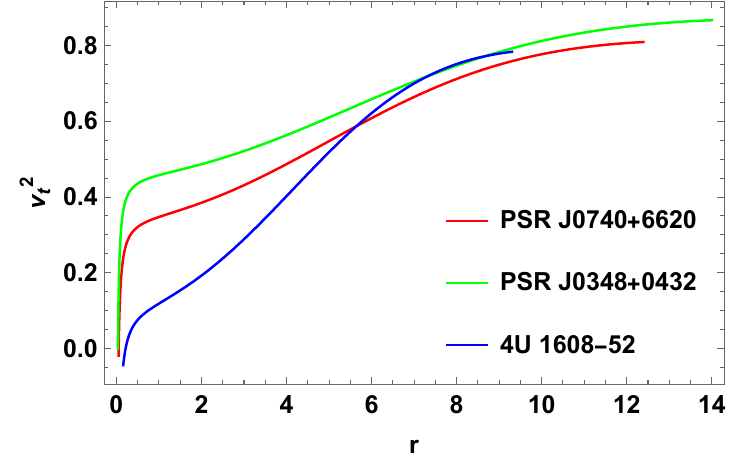}
    \caption{The plot of tangential speed of sound against $r$.}
    \label{fig:11}
\end{figure}

\begin{figure}
    \centering
    \includegraphics[width=0.75\linewidth]{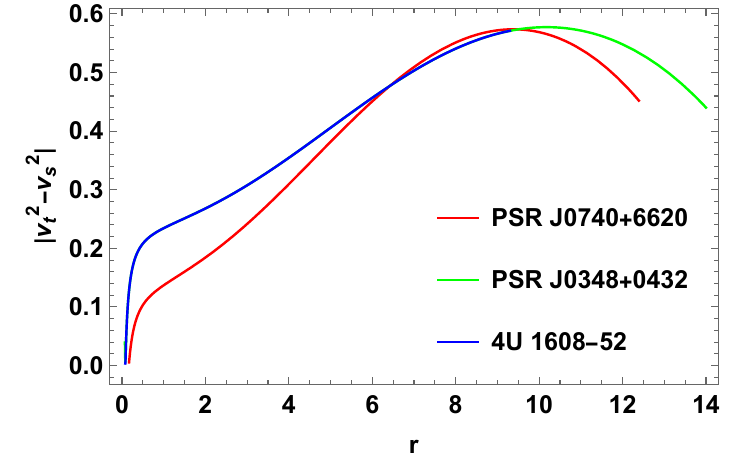}
    \caption{The plot of $\mid v_t^2-v_r^2\mid$ against $r$.}
    \label{fig:12}
\end{figure}

In the stability analysis based on causality conditions, two key sound velocities are considered: the radial speed of sound (\( v_r^2 \)) and the tangential speed of sound (\( v_t^2 \)). The causality condition sets an absolute upper limit, requiring that both velocities satisfy \( v_r^2 \leq 1 \) and \( v_t^2 \leq 1 \). Additionally, thermodynamic stability necessitates that \( v_r^2 \) and \( v_t^2 \) remain positive, ensuring that the conditions \( 0 \leq v_r^2 < 1 \) and \( 0 \leq v_t^2 < 1 \) hold throughout the stellar structure. The variations of \( v_r^2 \) and \( v_t^2 \) are graphically illustrated in Figs. (\ref{fig:10}), (\ref{fig:11}), and (\ref{fig:12}).

Herrera \cite{1992Herrera} introduced the cracking method to examine the stability of stellar models. Building on this approach, Abreu et al. \cite{2007Abreu} proposed a criterion that uses \( v_r^2 \) and \( v_t^2 \) to assess the stability of anisotropic star models. Fig. (\ref{fig:12}) shows that the stellar structure obeys the following condition 
\[0<\mid v_t^2-v_r^2\mid<1,\]
in $f(R)$ theory of gravity. 
\FloatBarrier
\FloatBarrier
\subsection{Energy conditions}
The next investigation will determine whether the energy conditions  are satisfied by the theoretical models \cite{1988Kolassis}-\cite{1984Wald}. These conditions represent essential significant events in modified theory of $f(R)$, determining the allowed energy and pressure distributions throughout spacetime. The necessary conditions consist of the null, weak, strong, and dominant energy conditions.  Compliance with these conditions supports the models physical validity by proving their compatibility with fundamental gravity and energy concepts. This evaluation is critical for establishing the theoretical validity and observational relevance of the findings in context of astrophysical and cosmological research. The energy conditions for current stellar configuration are fully satisfied. Both the interior and the surface of a compact star model must meet energy conditions in order to be considered physically realistic \cite{2021Brassel}-\cite{2021Maharaj}. Energy conditions are given as
\begin{itemize}
    \item $Null~~energy~~ condition~~~\Leftrightarrow~~~\rho^{eff}+p_{i}^{eff}\geq0,$
    \item $Weak~~energy~~condition~~~\Leftrightarrow~~\rho ^{eff}\geq 0,~\rho ^{eff}+p_{i}^{eff}\geq 0,$
    \item $Strong~~energy~~condition~~~~\Leftrightarrow~~~\rho ^{eff}+3p_{r}^{eff}+2p_{t}^{eff}\geq 0,~\rho ^{eff}+p_{i}^{eff}\geq 0,$
    \item $Dominant~~energy~~condition~~~\Leftrightarrow~~~\rho ^{eff}\geq 0,~\rho ^{eff}\pm p_{i}^{eff}\geq 0,$
\end{itemize}
\subsection{Gravitational redshift}
Gravitational redshift is the process where light or electromagnetic radiation emitted from a source in a strong gravitational field shifts to longer wavelengths as it escapes the field. This shift causes the light to move toward the red end of the spectrum. Modified theory of $f(R)$ predicts this effect, which occurs because gravity affects the energy of the escaping radiation. The degree of gravitational redshift $Z_s$ is given as

\begin{equation}
    Z_s=\frac{1}{\sqrt{1-\frac{2M}{r}}}-1
\end{equation}
 Bohmer and Harko \cite{bohmer2006bounds} demonstrated that for an anisotropic star, the surface redshift could reach a maximum of $Z_s\leq5$. Ivanov \cite{ivanov2002static} further refined this maximum value, establishing it at $Z_s=5.211$. Within this framework, the surface redshift for COs such as PSR J0740+6620, PSR J0348+04328, and 4U 1608-52 is constrained to $Z_s < 1$. This phenomenon becomes more pronounced near very massive objects like black holes or neutron stars, where the gravitational field is strong enough to noticeably stretch the wavelength of light. As light travels outward from such regions, its energy decreases, making it appear redder to observers in weaker gravitational fields. Furthermore, this redshift increases towards the object's boundary.
 \begin{figure}
 \centering
\includegraphics[width=0.75\linewidth]{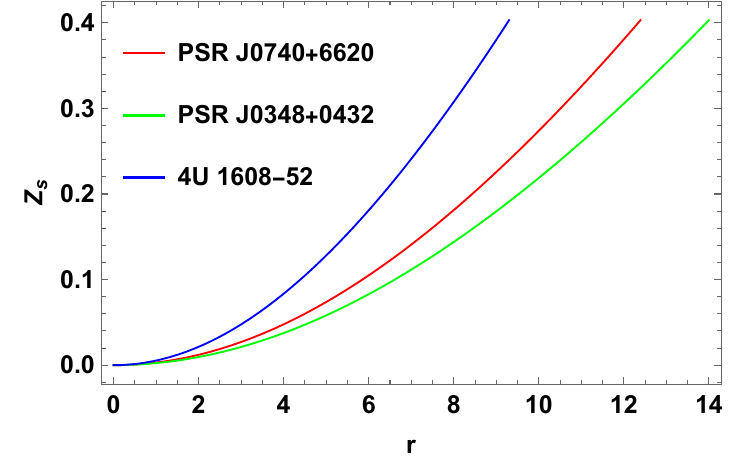}
\caption{The plot of gravitational redshift against $r$.}           \label{fig:13}                    
\end{figure}
\FloatBarrier
\subsection{Adiabatic index}
The adiabatic index parameter is an important mathematical tool to determine how pressure changes with density. In the study of neutron stars or the dynamics of gravitational collapse, the adiabatic index helps how ``stiff" the fluid is, affecting its stability against gravitational forces. According to Newtonian approximations, a stable compact structure requires an adiabatic index greater than $\gamma>4/3$  within the star interior. However, this criterion shifts for relativistic compact star models.
 The mathematical expression of adiabatic index is given as 
\begin{equation}
    \gamma=\Big(\frac{p_r+\rho}{p_r}\Big)\frac{dp_r}{d\rho}.
\end{equation}
\begin{figure}
\centering
\includegraphics[width=0.75\linewidth]{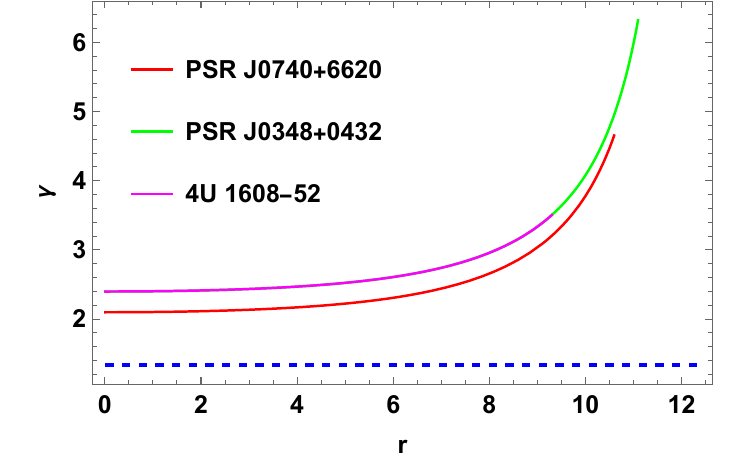}
\caption{Variation of adiabatic index against $r$.}
\label{fig:14}
\end{figure}
 \FloatBarrier
\section{Conclusion}
In this article, we used the theory of $f(R)$ to study the physical properties of a dense system in a spacetime that is spherically symmetric. To generate a more realistic set of solutions, we used the Buchdahl-I metric and Chaplygin EoS for metric functions with arbitrary constants. The $f(R)$ model established for PSR J0348+0432, PSR J0740+6620, and 4U 1608-52 provided the observational data that we used. Physical variables including $\rho$, $p_r$, $p_t$, and $\Delta$ were plotted along radial distance using observable and computed data. Additionally, the study examines the equilibrium of COs under gravitational, hydrostatic, and anisotropic forces, ultimately contributing to a deeper understanding of dense stellar objects such as PSR J0348+0432, PSR J0740+6620, and 4U 1608-52. Furthermore, the analysis of stability through the causality conditions, energy conditions, gravitational redshift, adiabatic index, and the cracking technique strengthens the accuracy of this model in $f(R)$ theory of gravity. 

Eqs. (\ref{13}), (\ref{15}), and (\ref{16}) provide mathematical expressions for $\rho$, $p_r$ and $p_t$ in $f(R)$ gravity illustrate how they depend on the radial coordinate and other parameters.
The analysis shows that $\rho$, $p_r$, and $p_t$ all exhibit positive values, which is necessary for the physical stability of the model. These quantities reach their maximum values at the center of the CO, particularly for parameter $H=0.3$. In Figs. (\ref{fig:1}), (\ref{fig:2}) and (\ref{fig:3}) show that as they move away from the center toward the surface of the object, these values gradually decrease. This behavior is typical for stable stellar configurations, where the central conditions are more extreme than on the outer surface. 
Anisotropy is crucial to assessing the stability of a CO. It indicates the difference between $p_r$ and $p_t$, and as shown in Fig. (\ref{fig:4}), the anisotropic pressure is highest at the center. However, as one moves outward, anisotropy decreases, leading to a situation where $p_t$ becomes less than $p_r$, creating an attractive force. This force counteracts gravitational attraction and is vital for the stability of the stellar object, preventing it from collapsing under its own gravity. Thus, the interaction between these pressures and the anisotropic force is fundamental to understanding the equilibrium and stability of COs in the modified $f(R)$ theory of gravity.

The Chaplygin gas EoS presents a compelling framework for modeling exotic matter that is hypothesized to exist within the interiors of COs, particularly those with higher masses. Given the crucial role of the EoS in characterizing the properties of such objects, it is plausible to conclude that the interiors of high-mass pulsars may be described by a Chaplygin gas EoS \cite{bhattacharjee2024maximum}. In our study, we have analyzed the fundamental characteristics of COs, such as the $\rho$, $p_r$, $p_t$, and $\Delta$, for the specific cases of 4U 1608-52, PSR J0740+6620, and PSR J0348+0432. Our findings indicate that the resulting models are consistent with realistic astrophysical scenarios. The models' physical viability is further guaranteed by the fact that the energy and causality conditions are met throughout the star interiors. We employed Herrera's cracking condition, the generalized TOV equation, and the modification of the $\gamma$ to evaluate the stability of the suggested models. We conclude that the current model, which is based on the modified Chaplygin gas EoS, offers a reliable and appropriate explanation of the internal characteristics of exotic COs, especially in the high-mass regime. Furthermore, it provides a stable, generalized, and physically feasible description of anisotropic COs.

In $f(R)$ theory, the conditions $\frac{d\rho}{dr}<0$, $\frac{dp_r}{dr}<0$, and $\frac{dp_t}{dr}<0$ indicates that $\rho$, $p_r$ and $p_t$ decreases outward from the center of COs. A positive \( \frac{d\rho}{dr} \), implying an outward increase in density, would be physically unrealistic and likely lead to instability or collapse. Thus,  $\frac{d\rho}{dr}<0$, $\frac{dp_r}{dr}<0$, and $\frac{dp_t}{dr}<0$ are critical for preserving the stability and realistic nature of COs within $f(R)$ as presented in Figs. (\ref{fig:5})-(\ref{fig:7}). These conditions are essential for creating physically realistic models of stellar structures, as it reflects a stable matter distribution with a denser core compared to the outer regions. 


\textbf{Energy Conditions:}
In anisotropic fluid distribution, this model satisfies several energy conditions, including the null, weak, strong, and dominant. These energy conditions are crucial for ensuring that the CO model represents a physically meaningful and stable object. The energy conditions are analyzed for various pulsars, including those previously mentioned. The results, illustrated in Figs. (\ref{fig:subfig8a}–\ref{fig:subfig8e}), indicate that these conditions peak at the core and gradually diminish toward the surface.

\textbf{Hydrostatic Equilibrium }
The hydrostatic equilibrium condition, governed by the TOV equation, ensures that the CO is balanced under $F_g$, $F_h$, and $F_a$. Eq. (\ref{TOV*}) provides the framework for this equilibrium, and the results in Figs. (\ref{fig:subfig9a}–\ref{fig:subfig9c}) demonstrate that the model is indeed in equilibrium. This balance is crucial for maintaining the structural integrity of the CO.

\textbf{Causality Conditions and Cracking Technique:}
In the context of relativistic stellar models, \( v_r^2 \), \( v_t^2 \)  must satisfy the causality condition, which requires that the speed of sound does not exceed the speed of light. Figures (\ref{fig:10}) and (\ref{fig:11}) illustrate that this condition is met, confirming the physical viability of the model. Furthermore, for the specific case where \( H = 0.3 \), the results indicate that the causality condition is maintained throughout the stellar configuration, reinforcing the idea that the model is realistic and stable under this choice of parameters. To further assess the stability of the CO, the cracking technique is applied.  As demonstrated in Fig. (\ref{fig:12}), no singularities appear in the model, and it remains well behaved throughout the range, confirming its stability.
 
\textbf{Gravitational Redshift:}
The gravitational redshift, $Z_s$, is another important characteristic of COs, representing the extent to which light is shifted to longer wavelengths due to the intense gravitational field. The redshift depends on the compactness of the CO, which is defined by its mass-to-radius ratio. The CO's redshift is zero in its center and reaches its highest value near its surface, as depicted in Fig. (\ref{fig:13}). The redshift values for neutron stars, characterized by strong gravitational fields, can range from $0.1-0.4$, depending on the specific mass and radius of the object.

\textbf{Adiabatic Index:}
The stability of charged COs against radial perturbations can be assessed using the a$\Gamma$, which quantifies the response of pressure to variations in density. This parameter plays a crucial role in determining whether a stellar structure can withstand small radial disturbances without undergoing collapse. For a compact star to be dynamically stable, the TOV criterion dictates that $\Gamma$ must be greater than $\frac{4}{3}$. If $\Gamma$ falls below this threshold, the gravitational force dominates over the pressure support, leading to instability and potential gravitational collapse. In this study, the variation of $\Gamma$ for different charged CO configurations is presented in Fig. (\ref{fig:14}), which shows that condition $\Gamma>\frac{4}{3}$ is satisfied throughout the stellar structure. This confirms that the model remains stable under radial perturbations, ensuring its reliability as a physically feasible representation of charged COs.
\section*{Appendix}
\begin{eqnarray}\label{16}
p_t&=&\frac{1}{L^3 (L - r^2 \chi)^2 (1 + r^2 \chi)^{10}}\times(1 + L) \chi \Bigg\{ 4 r^3 \beta \chi^4 \Big(432 - 36 r + 648 r^2 \chi - 576 r^3 \chi \nonumber\\&+& 540 r^4 \chi^2 + 2787 r^5 \chi^2 + 270 r^6 \chi^3 + 5337 r^7 \chi^3 + 90 r^8 \chi^4 + 2845 r^9 \chi^4 + 18 r^{10} \chi^5 \nonumber\\&+& 963 r^{11} \chi^5 + 2 r^{12} \chi^6 + 144 r^{13} \chi^6 + 16 r^{15} \chi^7\Big)
+ L^4 \Bigg( 6 + 3 (-44 + 15 r^2 - 24 \beta) \chi \nonumber\\&+& 2 \Big(72 r + 74 r^4 + 1512 \beta - r^2 (119 + 180 \beta)\Big) \chi^2 
+ 2 r \Big(288 r^2 + 140 r^5 + r^3 (221 - 381 \beta) \nonumber\\&-& 864 \beta - 4848 r \beta\Big) \chi^3 
+4 r^3 \Big(237 r^2 + 84 r^5 - 648 \beta - 5415 r \beta - 5 r^3 (-76 + 45 \beta)\Big) \chi^4 
\nonumber\\&+& 2 r^5 \Big(426 r^2 +133 r^5 + r^3 (760 - 331 \beta) - 1080 \beta - 6530 r \beta\Big) \chi^5 
+ 2 r^7 \Big(236 r^2 + 70 r^5 \nonumber\\&+& r^3 (329 - 160 \beta)- 540 \beta - 2494 r \beta\Big) \chi^6 + 2 r^9 \Big(84 r^2 + 24 r^5 + r^3 (61 - 51 \beta) - 180 \beta \nonumber\\&-& 498 r \beta\Big) \chi^7+ 2 r^{11} \Big(18 r^2 + 5 r^5 + r^3 (6 - 10 \beta) - 36 \beta - 68 r \beta\Big) \chi^8 
+ r^{13} \Big(4 r^2 + r^5 - 8 \beta \nonumber\\&-& 8 r \beta - 2 r^3 \beta\Big) \chi^9 \Bigg)- 2 L^3 \chi \Bigg( 6 + 36 \beta - 1584 \beta \chi - 46 r^9 \chi^5 + 2 r^{17} \chi^9 + r^{20} \chi^9 + 4 r^{11} \chi^6 \nonumber\\&&(17 - 18 \beta \chi) + 2 r^{15} \chi^8 (7 - 2 \beta \chi)+ r^{18} \chi^8 (10 + \chi - 2 \beta \chi) - 24 r^{13} \chi^7 (-2 + \beta \chi) 
+ 504 r^3 \chi^2\nonumber\\&& (-1 + 6 \beta \chi) + 54 r^7 \chi^4 (-7 + 10 \beta \chi)+ 144 r \chi (-1 + 18 \beta \chi) + 12 r^5 \chi^3 (-55 + 162 \beta \chi) 
\nonumber\\&+& r^6 \chi^2 (148 + 835 \chi - 312 \beta \chi - 23998 \beta \chi^2)+r^8 \chi^3 (280 + 2331 \chi - 569 \beta \chi - 14782 \beta \chi^2) 
\nonumber\\&+& r^4 \chi (45 - 212 \chi + 21 \beta \chi - 3540 \beta \chi^2)+2 r^{10} \chi^4 (168 + 1108 \chi - 251 \beta \chi - 2886 \beta \chi^2) 
\nonumber\\&+& r^{12} \chi^5 (266 + 938 \chi - 269 \beta \chi - 1048 \beta \chi^2)+ r^{14} \chi^6 (140 + 183 \chi - 92 \beta \chi - 142 \beta \chi^2) 
\nonumber\\&+& r^{16} \chi^7 (48 + 25 \chi - 19 \beta \chi - 6 \beta \chi^2)+3 r^2 (2 - 49 \chi + 36 \beta \chi + 2984 \beta \chi^2) \Bigg)
\nonumber\\&+& L^2 \chi^2 \Bigg( 144 \beta - 8 r^{17} \chi^8 + r^{22} \chi^9 
+ 2 r^{18} \chi^7 (24 + 13 \chi - 8 \beta \chi) - 2 r^{20} \chi^8 (-5 + \beta \chi)
\nonumber\\&+& 4 r^{15} \chi^7 (-17 + 6 \beta \chi) - 288 r^3 \chi (-1 + 9 \beta \chi) 
+ 12 r^{13} \chi^6 (-25 + 16 \beta \chi)- 144 r (-1 + 36 \beta \chi) 
\nonumber\\&+& 36 r^7 \chi^3 (-29 + 90 \beta \chi) + 8 r^{11} \chi^5 (-97 + 108 \beta \chi)+12 r^5 \chi^2 (-17 + 108 \beta \chi) \nonumber\\&+& 16 r^9 \chi^4 (-77 + 135 \beta \chi) 
- 12 r^2 (1 - 12 \beta + 672 \beta \chi) + 2 r^8 \chi^2 (74 + 558 \chi + 519 \beta \chi - 12196 \beta \chi^2) 
\nonumber\\&+& 2 r^{10} \chi^3 (140 + 1935 \chi + 212 \beta \chi - 9504 \beta \chi^2)+2 r^{12} \chi^4 (168 + 1829 \chi - 11 \beta \chi - 4266 \beta \chi^2) 
\nonumber\\&-& 2 r^{16} \chi^6 (-70 - 108 \chi + 31 \beta \chi + 84 \beta \chi^2)- 2 r^{14} \chi^5 (-133 - 708 \chi + 58 \beta \chi + 658 \beta \chi^2) 
\nonumber\\&+& 6 r^4 (1 - 69 \chi + 108 \beta \chi + 6158 \beta \chi^2)+ r^6 \chi (45 - 724 \chi + 1164 \beta \chi + 32388 \beta \chi^2) \Bigg) 
\nonumber\\&-&2 L r \chi^3 \Bigg( 864 \beta - 72 r \beta + r^{11} \Big(700 + \beta (331 - 308 \chi)\Big) \chi^4+r^{15} \Big(239 + \beta (51 - 198 \chi)\Big) \chi^6 \nonumber\\&+& 2 r^{16} \chi^7 + 2 r^{17} \Big(17 + \beta (5 - 16 \chi)\Big) \chi^7 
+ r^{19} (4 + \beta) \chi^8 + 4 r^{10} \chi^4 (59 - 126 \beta \chi) 
\nonumber\\&+& 4 r^{12} \chi^5 (21 - 26 \beta \chi) - 6 r^{14} \chi^6 (-3 + 2 \beta \chi) 
- 72 r^2 (-1 + 18 \beta \chi) - 72 r^4 \chi (-4 + 39 \beta \chi)
\nonumber\\&-& 6 r^8 \chi^3 (-71 + 240 \beta \chi) - 6 r^6 \chi^2 (-79 + 450 \beta \chi) 
+ 2 r^9 \chi^3 \Big(112 + 75 \beta (3 + 34 \chi)\Big) 
\nonumber\\&+& r^{13} \chi^5 \Big(639 - 2 \beta (-80 + 561 \chi)\Big) 
- 6 r^3 \Big(1 + \beta (-6 + 672 \chi)\Big) 
+ 3 r^5 \chi \Big(-27 + \beta (60 + 5162 \chi)\Big) 
\nonumber\\
&+& r^7 \chi^2 \Big(-121 + 3 \beta (127 + 7966 \chi)\Big) \Bigg)
\Bigg\}.
\end{eqnarray}

\end{document}